\documentclass[11pt,twoside]{article}

\usepackage{asp2006}
\usepackage{graphicx}

\begin{document}
\title{The Impact of Mergers on the Survival and Abundance of Disk-Dominated Galaxies}
\author{Jun Koda, Milo\v s Milosavljevi\'c, and Paul R. Shapiro}
\affil{Department of Astronomy, Texas Cosmology Center, University of Texas, 1 University Station C1400, Austin, TX 78712.}

\begin{abstract}
We study the formation of disk-dominated galaxies in a $\Lambda$CDM
universe. Their existence is considered to be a challenge for the
$\Lambda$CDM cosmology, because galaxy mergers isotropize stellar
disks and trigger angular momentum transport in gas disks, thus
fostering the formation of central stellar spheroids.  Here, we
postulate that the formation of stellar spheroids from gas-rich disks
is controlled by two parameters that characterize galaxy mergers, the
mass ratio of merging dark matter halos, and the virial velocity of
the larger merging halo. We utilize merger histories generated from
realizations of the cosmological density field to calculate the
fraction of dark matter halos that have avoided spheroid formation,
and compare the derived statistics with the spheroid occupation
fractions in surveys of nearby galaxies. We find, for example, that
the survival rate of disk-dominated galaxies in $\Lambda$CDM is just
high enough to explain the observed fractional representation of
disk-dominated galaxies in the universe if the only mergers which lead
to central spheroid formation are those with mass ratios $M_2/M_1 >
0.3$ and virial velocities $V_{\rm vir,1} > 55 \textrm{ km
  s}^{-1}$. We discuss the physical origin of this criterion, and show
that the dependence of the disk-dominated fraction on galaxy mass
provides a further test of the merger hypothesis.
\end{abstract}

\section{Introduction}
Disk-dominated galaxies with little or no bulge are frequently cited
as a challenge to the $\Lambda$CDM cosmology
\cite[e.g.][]{Kautsch:06}.  In $\Lambda$CDM cosmology, galaxy-hosting
halos grow hierarchically by mergers.  Major mergers destroy disks and
create elliptical galaxies/classical bulges.  On the other hand,
pseudobulges are thought to be created by secular evolution, not by
major mergers \citep{Kormendy:04}.  We consider what the requirements
for bulge-forming mergers must be in order for the merger history of
galactic halos in $\Lambda$CDM to account statistically for the
observed fraction of galaxies today which are ``disk-dominated'' --
i.e., those with either no bulge or only a pseudobulge. [For additional details, see, \citet{Koda:07}.]

\section{A Simple Merger Hypothesis}
\label{sec:hypothesis}
We explore a simple model in which bulge formation due to merger is
controlled by two parameters, mass ratio $\mu= M_2/M_1 (\le 1)$ and
the virial velocity of the larger halo, $V_{\rm vir,1}$. The
hypothesis of the model is that a bulge forms if and only if,

\begin{equation}
  \label{eq:bulge-forming-criterion}
    \mu > \mu_{\rm crit} \textrm{ and } V_{\rm vir,1} > V_{\rm vir,crit}.
\end{equation}

\begin{figure}
  \begin{tabular}{cc}
    \begin{minipage}{0.45\textwidth}
      \begin{center}
	\includegraphics[height=6.0cm,clip]{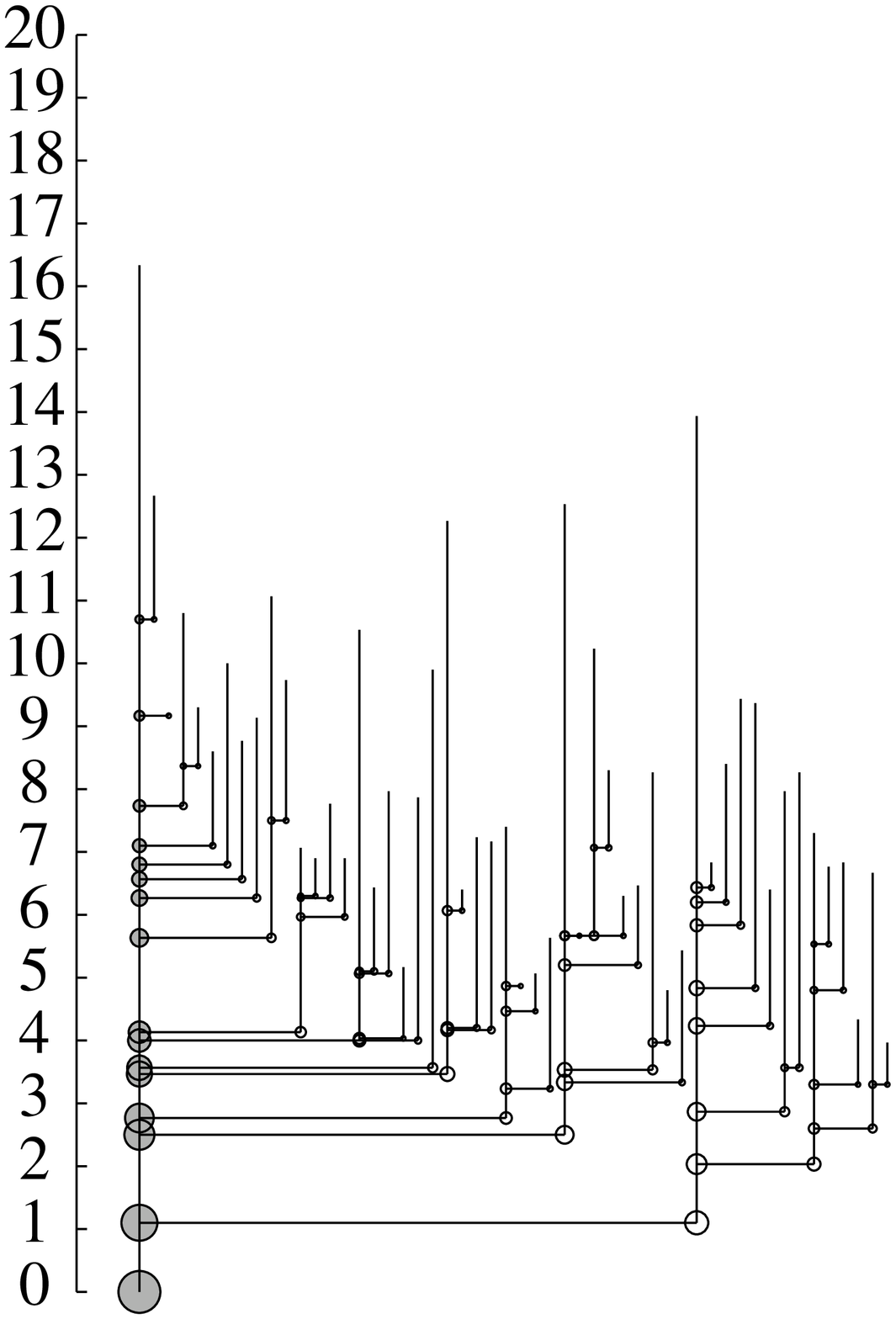}
      \end{center}
    \end{minipage}
    &
    \begin{minipage}{0.45\textwidth}
      \begin{center}
	\includegraphics[height=6.0cm,clip]{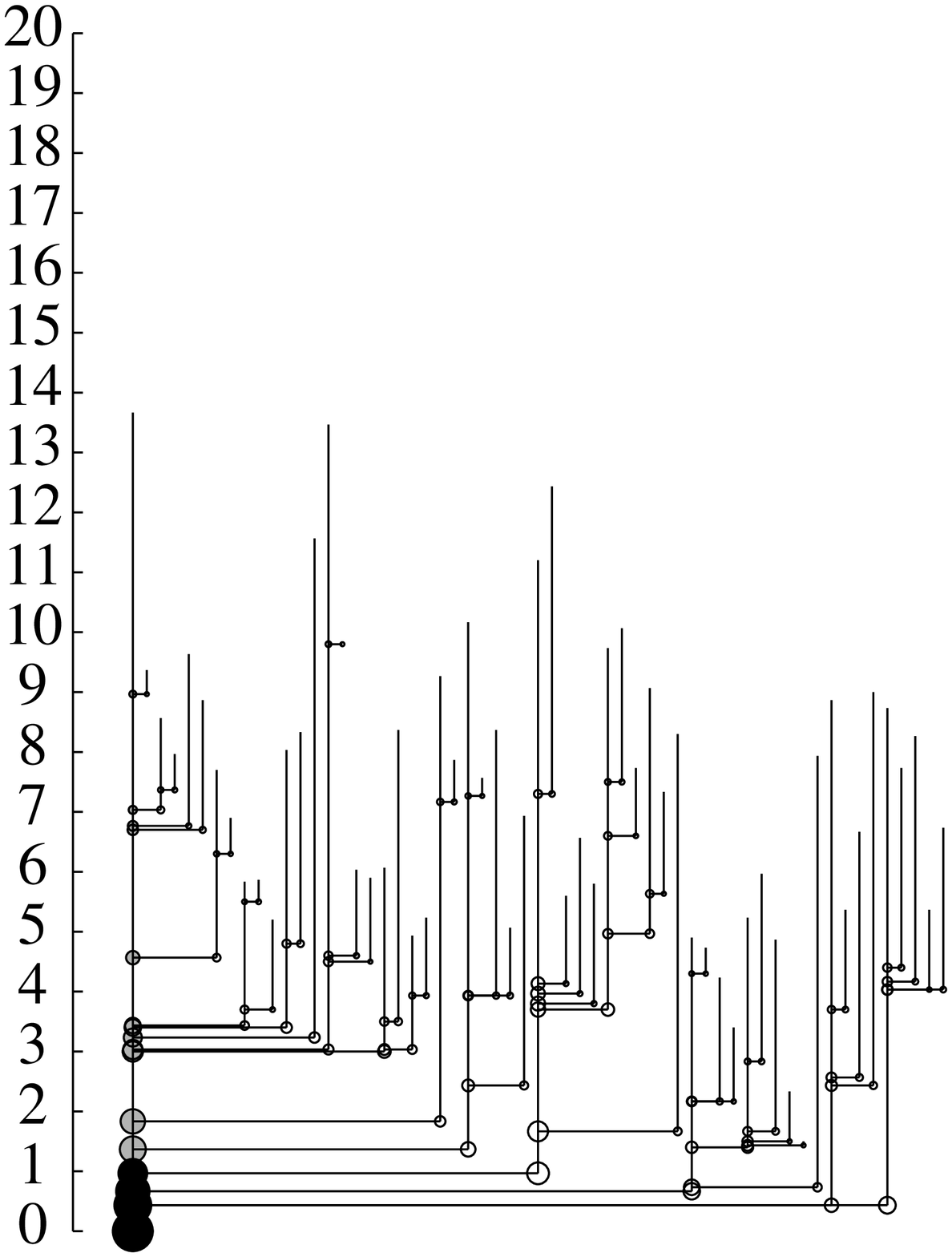}
      \end{center}
    \end{minipage}
  \end{tabular}

  \caption{Examples of merger trees. Gray circles represent
    disk-dominated galaxies and black circles represent galaxies with
    classical bulges, respectively, under a bulge-forming criterion
    $\mu_{\rm crit}=0.3$, $V_{\rm vir,crit} = 55 \textrm{ km
      s}^{-1}$. Radii of circles are proportional to
    $M^{1/3}$. Mergers with $\mu < 0.05$ are not plotted in the
    figure. Vertical axis label is redshift.}
  \label{fig:trees}
\end{figure}

\textbf{Motivation:} Mergers which make bulges are generally assumed
to require a merging-halo mass ratio, $\mu$, above some threshold,
$\mu_{\rm crit}$, consistent with $N$-body and some hydro simulations
\citep[e.g.,][]{Bournaud:05, Cox:07}.  We hypothesize here an
additional dependence on $V_{\rm vir,1}$, the virial velocity of the
larger halo. The virial velocity characterizes gas dynamical effects.
Bulge formation requires angular-momentum transport, which strong
shocks can promote. Shock-induced radiative cooling and compression
also promote gravitational instability. The strength of merger shocks
is characterized by the Mach number ${\cal M} \propto V_{\rm vir}$.
If the IGM was pre-heated by reionization, gas pressure can prevent gas from
collapsing into galaxies \citep[``Jeans-mass filtering,'' ][]{Shapiro:94} if
the virial velocity is too low. The threshold value of this Jeans-mass
filtering is uncertain \citep[$\sim 30 - 80 \textrm{ km s}^{-1}$,
  e.g.,][and references therein]{Okamoto:08}.  Mergers of pre-existing stellar systems with
$\mu > \mu_{\rm crit}$ still require $V_{\rm vir,1} > V_{\rm
  vir,crit}$ to make enough stars before merger (i.e., star formation
requires gaseous baryons but halos may not retain if $V_{\rm vir}$ too
low).\\

\textbf{Method: Merger Tree.}  We generate merger trees for the
galactic halos in a comoving volume 40 Mpc using a publicly available
code PINOCCHIO \citep{Monaco:02} based on Lagrangian perturbation
theory.  The code gives halo masses and their merger histories that
are in good agreement with $N$-body simulations, but with less
computation.  We follow the most massive progenitor of each halo, in
mass range $5 \times 10^{10} M_\odot < M < 10^{12} M_\odot$, and check
the bulge-forming condition (Eq. \ref{eq:bulge-forming-criterion}) for
all mergers (see Fig.~\ref{fig:trees}). We assume that a halo hosts a
disk-dominated galaxy if none of its mergers satisfies the bulge
formation criterion.  We vary $\mu_{\rm crit}$ and $V_{\rm vir,crit}$
to see which critical values are consistent with the observed fraction
of disk-dominated galaxies.

\section{Results}
\label{sec:results}

\begin{figure}
  \begin{tabular}{cl}
    \begin{minipage}{0.47\textwidth}
      \begin{center}
	\includegraphics[height=7.0cm,clip]{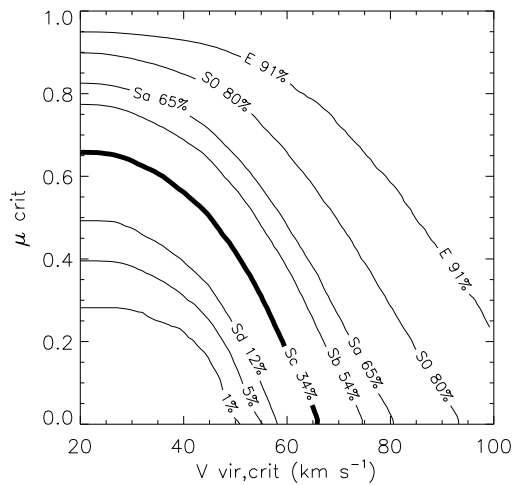}
      \end{center}
    \end{minipage}
    &
    \begin{minipage}{0.47\textwidth}
      \begin{center}
        \caption{The fraction of disk-dominated galaxies that results
          from bulge formation criteria characterized by the critical
          merger mass ratio, $\mu_{\rm crit}$, and the critical virial
          velocity of the larger halo at merger, $V_{\rm vir,crit}$.
          The points along the bold contour ``Sc 34\%'' (e.g.,
          $\mu_{\rm crit} = 0.3, V_{\rm vir,crit} = 55 \textrm{ km
            s}^{-1}$) constitutes our theoretical expectation for the
          bulge-forming threshold criteria that is compatible with
          observation (See
          \S~\ref{sec:results}). }
        \label{fig:disk-dominated-fraction}
      \end{center}
    \end{minipage}
  \end{tabular}
\end{figure}

Figure~\ref{fig:disk-dominated-fraction} shows the fraction of
galaxies without classical bulge (disk-dominated galaxies) as a
function of the bulge formation criterion $(\mu_{\rm crit}, V_{\rm
  vir,crit})$, which is the result of the model described in
\S~\ref{sec:hypothesis} Some of the contours are also labeled by the
morphological type of galaxy in the {\it Tully Catalogue}
\footnote{http://haydenplanetarium.org/universe/duguide/exgg\_tully.php}
such that the observed fraction of galaxies of that morphological type
or later equals the theoretical fraction of disk-dominated galaxies.
We assume that galaxies of type Sc and later are the disk-dominated
population. Fraction of Sc or later is $34\%$ in the Tully subsample, which
is also consistent with other observations \citep[e.g.,][]{Kautsch:06,
  Barazza:07}. So, the $(\mu_{\rm crit}, V_{\rm vir,crit})$-values for
which our bulge formation hypothesis predicts the observed
disk-dominated-galaxy fraction $34\%$ are shown by the bold contour in
Fig.~\ref{fig:disk-dominated-fraction}, labelled ``Sc 34\%,'' e.g., 
$\mu_{\rm crit} = 0.3, V_{\rm vir,crit} = 55 \textrm{ km s}^{-1}$. 
Somewhat smaller $\mu_{\rm crit}$ and larger $V_{\rm vir,crit}$, or
vice versa, are also plausible. However, $V_{\rm vir,crit}$ cannot be
larger than $65 \textrm{ km s}^{-1}$ because it will overproduce
disk-dominated galaxies.

\begin{figure}
  \begin{center}
    \includegraphics[height=6.0cm,clip]{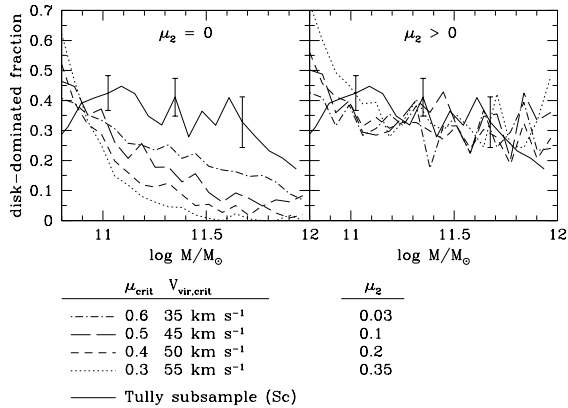}
  \end{center}
  \caption{Disk-dominated fraction as a function of galaxy-hosting
    halo mass for the two-parameter model $(\mu_{\rm crit}, V_{\rm
      crit})$ ({\it left panel}) and the three-parameter model with
    additional parameter $\mu_2$ ({\it right panel}), in which we
    assume that a merger remnant would not be observed as a classical
    bulge if its mass were smaller than $\mu_2$ times present halo
    mass. Solid line is the observed fraction of Sc and later in the
    Tully subsample. Our model with only two parameters ({\it left})
    does not agree with the observed mass dependence, but the
    agreement becomes reasonably good by introducing the third
    parameter $\mu_2$ ({\it right}).}
  \label{fig:mass-dependence}
\end{figure}

\section{Conclusion} In order to explain the observed fraction of
disk-dominated galaxies within $\Lambda$CDM cosmology, we propose a
bulge-forming criterion such that only those mergers with $M_2/M_1 >
\mu_{\rm crit} \sim 0.3$ and $V_{\rm vir,1} > V_{\rm vir, crit} \sim 55 \textrm{ km s}^{-1}$
form bulges.  The validity of this bulge formation criterion needs to
confirmed by further study of low-mass mergers $(V_{\rm vir,1} \sim 30
- 60 \textrm{ km s}^{-1})$ by, e.g., hydro simulations. The mass
dependence of the disk-dominated fraction (Fig.~\ref{fig:mass-dependence})
can be explained, too, if merger remnants which are small 
compared to the present halo mass are too small to be observed as a
classical bulge.  Future surveys of bulgeless/pseudobulged galaxy
fraction as a function of mass would be useful for determining
the bulge-formation criterion and testing this hypothesis.

\acknowledgements 
We would like to thank Shardha Jogee for detailed comments,
and John Kormendy for inspiring and illuminating discussions.  This
work was supported in part by NSF grant AST-0708795 to M.~M., and NASA
ATP grants NNG04G177G, NNX07AH09G, SAO TM8-9009X, and NSF grant
AST-0708176 to P.~R.~S.

\end{document}